\documentstyle[aps,epsf,bezier,twocolumn]{revtex}

\newcommand{\bqn}{\begin{eqnarray}}
\newcommand{\eqn}{\end{eqnarray}}
\newcommand{\beq}{\begin{equation}}
\newcommand{\eeq}{\end{equation}}

\def\bk{{\mbox{\boldmath$k$}}}

\def\bke{{\mbox{\boldmath$k$}_{e}}}

\def\bp{{\mbox{\boldmath$p$}}}

\begin{document}

\title{Can we learn anything new from  polarization
  observables of the deuteron disintegration near threshold?}

\author{S.G. Bondarenko, V.V. Burov}
\address{Bogoliubov Laboratory of Theoretical Physics,
JINR Dubna, 141980 Russia}
\author{M. Beyer}
\address{Max Planck AG ``Theoretical Many Particle Physics'', Rostock
University, 18051 Rostock, Germany}
\author{S.M. Dorkin}
\address{Far Eastern State University Vladivostok, 690000 Russia}

\maketitle

\begin{abstract}
  We discuss polarization characteristics of the deuteron
  disintegration near the threshold energy. Due to the small relative
  energy of the outgoing $np$ pair, the dominant amplitude for this
  process is a $1^+\rightarrow 0^+$ transition. Relativistic
  covariance requires only one electromagnetic transition form factor
  for this amplitude.  Subsequently this leads to substantial
  simplifications in the formulas of the polarization observables and
  allows to draw conclusions independent of the details of the
  interaction.
\end{abstract}


\section*{}
Electrodisintegration of the deuteron near the threshold
energy~\cite{sacle,slac} is a rich source of information on the
deuteron structure and the nucleon--nucleon interaction. Due to the
small value of the relative energy of the $np$--pair (in the rest
frame) and because of kinematical reasons (backward electron
scattering), it is possible to suppose that the $^1S_0$--component
dominates the final state~\cite{leide}.  Commonly, this reaction is
described in a nonrelativistic framework with non--nucleon degrees of
freedom (meson exchange currents~\cite{mathiot,foldy},
$\Delta$--isobars~\cite{mathiot}, $N^{*}$-excitations~\cite{glozman})
and relativistic corrections~\cite{burov,tamura} to improve the
agreement with experimental data.  These corrections are important
because the nonrelativistic impulse approximation has a zero in the
double differential cross--section
$d^2\sigma/dE_e^{\prime}d\Omega_e^{\prime}$ at $-q^2 \sim 40$fm$^{-2}$
which is not observed in the experimental data~\cite{sacle,slac,will}.
In turn this reaction is a strong support for non--nucleonic degrees
of freedom and a good paradigm to study them.  The results of the
calculations have shown a strong sensitivity to the nucleon
electromagnetic form factors, the way meson exchange currents are
constructed, as well as the way $N^{*}$- and $\Delta$-states are
included into the calculations.  Relativistic approaches to describe
the electrodisintegration of deuterons near the threshold energy have
also been developed, namely the light-front approach~\cite{karmanov},
and the Bethe--Salpeter approach~\cite{bbbd:dis}.

The general form of the deuteron electrodisintegration amplitude
$M_{fi}$ utilizing the one photon approximation
may be written in the following way:
\begin{eqnarray}
M_{fi}=ie^2{\bar u(k_e^{\prime},s_e^{\prime})}
\gamma^{\mu}u(k_e,s_e)\;\frac{1}{q^2}
\;\langle np | j_{\mu} | D {\cal M} \rangle,
\label{eqn:M}
\end{eqnarray}
where $u(k_e,s_e)$ denotes the free electron spinor with 4-momentum
$k_e$ and spin $s_e$, and $q=k_e-k_e^{\prime}$ is the 4-momentum
transfer.  The hadronic transition matrix element $\langle np |
j_{\mu} | D {\cal M} \rangle$ is from the deuteron state $| D {\cal M}
\rangle$ with 4-momentum $K$ and total angular momentum projection
$\cal M$ to the final $np$ state with 4-momentum $P=K+q$, where
$j_\mu$ is the electromagnetic current operator.

In the $^1S_0$ approximation, the covariant form of the matrix
hadronic matrix element is due to a $1^+\rightarrow 0^+$ transition.
It depends on one scalar function only (because of parity
conservation), which is the electromagnetic transition form factor
$V(s,q^2)$. This form factor contains all the structure information,
viz. the deuteron and $np$--pair wave functions. It is defined via
\begin{eqnarray}
\langle np (^1S_0) | j_{\mu} | D {\cal M} \rangle &=&
i\epsilon_{\mu\alpha\beta\gamma}\;\xi^{\alpha}_{\cal M}\;
q^{\beta}\;K^{\gamma}\;V(s,q^2)
\nonumber\\
&\equiv&\xi^{\alpha}_{\cal M}\;G_{\mu\alpha}\;V(s,q^2),
\label{formc1}
\end{eqnarray}
where $s=P^2$, and the deuteron polarization 4-vector
$\xi^{\alpha}_{\cal M}$ has been introduced.

Note, that this particular simple form of the amplitude is only valid
if the $^1S_0$--channel dominates the final state. In other words, the
reaction is dominated by the $M1$ transition. The inclusion of other
$NN$ final channels (e.g. $^3S_1-^3D_1$, $^1P_1$, $\dots$) would lead
to more complicated expressions and to additional form factors in
eq.(\ref{formc1}).

The purpose of the present letter is to give unique observables that
enable us to  check experimentally the assumptions that lead to
eq.(\ref{formc1}). Some polarization observables  turn
out to be independent of the form factor $V(s,q^2)$. This will be
shown below.

Using eq.~(\ref{eqn:M}) the differential cross-section may be obtained
in the standard way, see ref.~\cite{bjorken}.  Here it is useful to
introduce leptonic $l_{\mu\nu}$ and hadronic $W^{\mu\nu}$ tensors.
The differential cross section then reads:
\begin{eqnarray}
\frac{d^{2}\sigma}{dE_e^{\prime}d\Omega_e^{\prime}}\; =\;
\frac{\alpha^2}{q^4}\; \frac{|\bke^{\prime}|}{|\bke|}\;
l_{\mu\nu}\; W^{\mu\nu},
\end{eqnarray}
with $\alpha=e^2/4\pi$.
The leptonic tensor is given by
\begin{equation}
l_{\mu\nu} = 2 (k_{e\mu}k^{\prime}_{e\nu}+k^{\prime}_{e\mu}k_{e\nu})+
q^2 g_{\mu\nu}
+2im_{e}{\epsilon}_{\mu\nu\alpha\beta}q^{\alpha}s_{e}^{\beta}.
\end{equation}
The hadronic tensor $W_{\mu\nu}$ has the following form:
\begin{eqnarray}
&&W^{\mu\nu} =
\langle np (^1S_0) | j^{\mu} | D {\cal M} \rangle
\langle D {\cal M} | j^{\dagger\nu} | np (^1S_0)\rangle
\label{W:def}\\
&&
\times\frac{(2\pi)^3}{2M}
\int \delta(K+q-k_p-k_n) \frac{d{\bk}_p}{2E_{p}(2\pi)^3}
\frac{d{\bk}_n}{2E_{n}(2\pi)^3}.
\nonumber\end{eqnarray}
Using the general form of the hadronic transition current
eq.(\ref{formc1}), the hadronic tensor may be written as
\begin{equation}
W^{\mu\nu} = R\;G^{\mu\alpha}\;\rho_{\alpha\beta}\;
G^{*\nu\beta}\;V^2(s,q^2),
\label{eqn:W2}
\end{equation}
where $R$ is a purely kinematical factor. It reads
\begin{equation}
R=\frac{1}{8\pi^2} \frac{|\bp^{*}|}{\sqrt{s}},
\qquad |\bp^{*}|=\sqrt{\frac{s}{4}-m^2}.
\end{equation}
In eq.~(\ref{eqn:W2}) the density matrix
$\rho_{\alpha\beta}$ of the
deuteron is given by
\begin{eqnarray}
\rho_{\alpha\beta} &=& \frac{1}{3}(-g_{\alpha\beta}+
\frac{K_{\alpha}K_{\beta}}{M^2})
+\frac{1}{2M}i
{\epsilon}_{\alpha\beta\gamma\delta} K^{\gamma} s_{D}^{\delta}
\nonumber\\
&&-\bigl[\frac{1}{2}
\bigl(
{(W_{\lambda_1})}_{\alpha\rho}
{(W_{\lambda_2})}^{\rho}_{~\beta}+
{(W_{\lambda_2})}_{\alpha\rho}
{(W_{\lambda_1})}^{\rho}_{~\beta}
\bigr)
\\
&&- \frac{2}{3}
(-g_{\lambda_1\lambda_2}+\frac{K_{\lambda_1}K_{\lambda_2}}{M^2})
(-g_{\alpha\beta}+\frac{K_{\alpha}K_{\beta}}{M^2}) \bigr] 
p_{D}^{\lambda_1
\lambda_2},
\nonumber\end{eqnarray}
where
${(W_{\lambda})}_{\alpha\beta}=
i\epsilon_{\alpha\beta\gamma\lambda}K^{\gamma}/M$,
$s_{D}$ is the spin vector and $p_{D}$ is the alignment tensor
of the deuteron.  Using this explicit form of the density
matrix, the hadronic tensor may be written as
(the electron mass is neglected)
\begin{eqnarray}
W^{(u)}_{\mu\nu}&=&\frac{1}{3}\;R\;
\bigl[g_{\mu\nu}(q^2M^2-(Kq)^2)\nonumber\\
&&+(K_{\mu}q_{\nu}+q_{\mu}K_{\nu})(Kq)\nonumber\\
&&-K_{\mu}K_{\nu}q^2-q_{\mu}q_{\nu}M^2\bigr]\;V^2(s,q^2),
\nonumber\\
W^{(v)}_{\mu\nu}&=&\frac{1}{2}\;R\;M\;(s_Dq)\;
i {\epsilon}_{\mu\nu\alpha\beta}\;q^{\alpha}\;K^{\beta}\;V^2(s,q^2),
\label{tensor}\\
W^{(t)}_{\mu\nu}&=&R\;\Bigl[\frac{1}{2}
[\epsilon_{\mu\lambda_2\alpha\beta}
\epsilon_{\lambda_2\nu\gamma\delta}
+\epsilon_{\mu\lambda_2\alpha\beta}
\epsilon_{\lambda_1\nu\gamma\delta}]
K^{\alpha}K^{\gamma}q^{\beta}q^{\delta}
\nonumber\\
&&+\frac{1}{3}\bigl(-g_{\lambda_1\lambda_2}
+\frac{K_{\lambda_1}K_{\lambda_2}}
{M^2}\bigr)\bigl[g_{\mu\nu}(q^2M^2-(Kq)^2)
\nonumber\\
&&+(K_{\mu}q_{\nu}+q_{\mu}K_{\nu})(Kq)-K_{\mu}K_{\nu}q^2
\nonumber\\
&&-q_{\mu}q_{\nu}M^2\bigr]\;p_D^{\lambda_1\lambda_2}
\Bigr]\;V^2(s,q^2).
\nonumber\end{eqnarray}
The superscripts ${(u,v,t)}$ denote unpolarized, vector polarized and
tensor polarized cases, respectively.

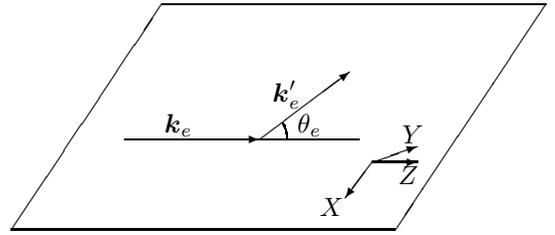
\begin{figure}[h]
%\unitlength=1mm
\unitlength=0.6mm
\special{em:linewidth 0.4pt}
\linethickness{0.4pt}
\begin{picture}(100.00,90.00)

%%%%%%%%%%%%%%%%%%%%%%%%%%%%%%%%%%%%%%%%%%%%%%%%
%% Lines for the plane
%%%%%%%%%%%%%%%%%%%%%%%%%%%%%%%%%%%%%%%%%%%%%%%%

\put(48.50,90.00){\line(1,0){85.00}}
\put(100.00,40.00){\line(-1,0){85.00}}

\put(133.50,90.00){\line(-2,-3){33.40}}
\put(15.00,40.00){\line(2,3){33.40}}

%%%%%%%%%%%%%%%%%%%%%%%%%%%%%%%%%%%%%%%%%%%%%%%%

% initial electron
\put(40.00,60.00){\vector(1,0){30.00}}
% final electron
\put(70.00,60.00){\vector(4,3){20.00}}
% line along direction of the initial electron
\put(70.00,60.00){\line(1,0){22.00}}

% axe Z
\put(95.00,55.00){\vector(1,0){10.00}}
% axe X
\put(95.00,55.00){\vector(-3,-4){6.00}}
% axe Z
\put(95.00,55.00){\vector(3,1){10.00}}

% angle theta_e
\bezier{32}(75.00,63.50)(76.00,62.00)(76.00,60.00)

%%%%%%%%%%%%%%%%%%%%%%%%%%%%%%%%%%%%%%%%%%%%%%%%%%%
%% Notations
%%%%%%%%%%%%%%%%%%%%%%%%%%%%%%%%%%%%%%%%%%%%%%%%%%%
\put(81.00,63.00){\makebox(0,0)[cc]{$\theta_e$}}
\put(52.00,63.00){\makebox(0,0)[cc]{$\bk_e$}}
\put(76.00,70.00){\makebox(0,0)[cc]{$\bk_e^{\prime}$}}
\put(86.00,45.00){\makebox(0,0)[cc]{$X$}}
\put(104.00,61.00){\makebox(0,0)[cc]{$Y$}}
\put(103.00,52.00){\makebox(0,0)[cc]{$Z$}}
\end{picture}       
\protect\caption{Kinematical situation for $ed\to enp$ in the
  laboratory system. The electron
momenta are in the $XZ$ plane, the $Y$ axis is directed
to form a right handed system.}
\label{fig:kinem}
\end{figure}

With the general form of hadronic tensor $W^{\mu\nu}$
eq.~(\ref{tensor}) it is straight forward to calculate asymmetries of
the deuteron disintegration near threshold. First we consider the spin
correlation of the incident particles, e.g.
\begin{eqnarray}
A=\frac {d\sigma(\uparrow,D)-d\sigma(\downarrow,D)}
 {d\sigma(\uparrow,D)+d\sigma(\downarrow,D)},
\label{assym}
\end{eqnarray}
where $d\sigma$ is the differential cross section,
$\uparrow(\downarrow)$ denotes the helicity $\lambda_e=+1(-1)$ of the
incoming electron and $D$ the polarization state of the deuteron,
which might be vector or tensor type. We assume the initial electron
moving along the $Z$ axis, and $\theta_e$ is the electron scattering
angle. The scattering plane of the electron is in the $XZ$ plane (see
Fig.~\ref{fig:kinem}).  Then the vectors $k_e$ and $k_e^{\prime}$ obtain
the following form:
\begin{eqnarray}
  k_e&=&(E_e,0,0,E_e),\nonumber\\
  k_e^{\prime}&=&
(E_e^{\prime},-E_e^{\prime}\sin{\theta_e},0,E_e^{\prime}\cos{\theta_e}).
\end{eqnarray}

First we consider the case of {\em vector} polarized
deuterons. If the direction of the deuteron polarization is
parallel to the $Z$--axis, then the correlation is given by
\begin{eqnarray}
A_{\parallel} &=& \frac{3}{2}\ \kappa
\frac{(E+E^{\prime})
(E-E^{\prime}\cos{\theta_e})}{(E+E^{\prime})^2
-2EE^{\prime}\cos^{2}{\theta_e/2}},
\label{A:parallel}
\end{eqnarray}
where $\kappa$ is the degree of polarization of the deuterons. Note,
that the dependence on the form factor $V(s,q^2)$ disappears.  For the
case of the backward scattering ($\theta_e=180^{\circ}$)
eq.~(\ref{A:parallel}) even simplifies to
\begin{equation}
A_{\parallel} = \frac{3}{2}\ \kappa.
\end{equation}
If the polarization of the deuteron is parallel to the $X$--axis, then
\begin{eqnarray}
A_{\perp} = \frac{3}{2}\ \kappa 
\frac{(E+E^{\prime})E^{\prime}\sin{\theta_e}}
{(E+E^{\prime})^2-2EE^{\prime}\cos^{2}{\theta_e/2}}.
\label{A:perp}
\end{eqnarray}

These formulae may be generalized to arbitrary polarization direction
of the deuteron  given by the angles $(\vartheta,\varphi)$, viz.
\begin{eqnarray}
\lefteqn{A(\vartheta,\varphi) = \frac{3}{2}\ \kappa(E+E^{\prime})}
\label{A:arb}\\&&
\times\frac{(E^{\prime}\sin{\theta_e}\sin{\vartheta}\cos{\varphi}+
(E-E^{\prime})\cos{\theta_e}\cos{\vartheta})}
{(E+E^{\prime})^2-2EE^{\prime}\cos^{2}{\theta_e/2}}.
\nonumber\end{eqnarray}

Now, consider the case of {\em tensor} polarization of the inital
target. If the initial deuteron is only aligned due to a $p_{D\,zz}$
component, then the cross section reads
\begin{eqnarray}
&&d\sigma(p_{zz})=d\sigma[1 + A_{zz} p_{D\,zz}],
\label{ratio}\\
&& A_{zz}=\frac{4E_e^2+4E_e^{\prime\;2}-4E_eE_e^{\prime}\cos{\theta_e}+
3E_e^{\prime\;2}\cos{2\theta_e}}
{4(E_e+E_e^{\prime})^2-2E_eE_e^{\prime}\cos^2{\theta_e/2}},
\nonumber\end{eqnarray}
where $A_{zz}$ is the tensor analyzing power. 
For the backward scattering the analyzing power is
\begin{eqnarray}
A_{zz} = 1.
\label{rback}\end{eqnarray}

As a last example consider the scattering of polarized initial
electrons from unpolarized deuterons. Then the polarization transfer is
maximal and the polarization of the final electrons coincides with the
polarization of initial electron beam.  Note, that the same holds
for the scattering of electrons from a structureless target
(Coloumb scattering).

We have shown that it is possible to obtain rather simple expressions
for the polarization observables in the reactions
$\vec{e}(\vec{d},np)e$ and $\vec{e}(d,np)\vec{e}$. These simple
relations given in eqs.~(\ref{A:parallel}), (\ref{A:perp}), and
(\ref{ratio}) fully demonstrate the advantages of a covariant
formalism and gives a positive answer to our initial question.  In a
nonrelativistic treatment usually utilized for disintegration at the
threshold energy these relations are not so obvious. This is due to
different corrections that have to be taken into account in the
transition matrix elements. In fact, it is possible to check the
assumption of the $^1S_0$ dominance in the final channel
experimentally. Any deviation of the structure independent values
given above from the experimental data can only be due to other
components in the final state. Besides the $^1S_0$ dominance the
relations are based on the general form of the transition amplitude,
which is constructed using invariance principles only.  It is
important to note that the form factor $V(s,q^2)$ is a very complex
function containing all the information of the $NN$--interaction,
electromagnetic properties of the nucleons, etc.  Nontheless, this
function vanishes in the final relations.

 On the other hand, if no accidental
cancellations occur between higher partial components,
eqs.(\ref{A:parallel}) and (\ref{ratio})
may be used to calibrate the deuteron target, i.e. to determine the
polarization degree of the deuteron target.

We wish to thank Karmanov V.A., Shapiro I.S., Kolybasov V.M., Kaptari
L.P.  and Titov A.I. for helpful discussions. This work has been
supported by the Heisenberg--Landau program of the Deutsche
Forschungsgemeinschaft.

\vskip 30mm

\end{document}